\documentstyle[12pt]{article}
\global\arraycolsep=2pt

\newcommand{\be}{\begin{eqnarray}}
\newcommand{\ee}{\end{eqnarray}}
\newcommand{\la}{\langle}
\newcommand{\ra}{\rangle}
\begin{document}

\begin{titlepage}
\begin{flushright}
SMU-HEP-93-23\\
hep-ph/9311349\\
November 1993
\end{flushright}

\vspace{0.3cm}
\begin{center}
\Large\bf What do   heavy-light ($ Q\bar{q} $) quark systems
tell us about \\
 QCD vacuum properties?
 \end{center}

\vspace {2cm}

 \begin{center} {\bf Ariel R. Zhitnitsky\footnote{
On leave of absence from Budker Institute of Nuclear Physics,\\
Novosibirsk,630090,Russia.\\
e-mail addresses:arz@mail.physics.smu.edu, ariel@sscvx1.ssc.gov}}
 \end{center}

\begin{center}
{\it Physics Department, SMU , Dallas, Texas, 75275-0175}

\end{center}
\begin{abstract}

Arguments in favor of a large magnitude ( at least
two- three times bigger
than its factorized value)  of the mixed vacuum condensate
$ \la \bar{q}G_{\mu\nu}^a G_{\mu\nu}^a q \ra$  are given. The analysis is
based on  the strict inequalities  which
follow from the QCD sum rules  method and  on   very plausible
phenomenological assumptions  like $ m_{B_s}  >  m_{B_u} $
for the few lowest
exited  $ Q\bar{q} $ states in a heavy quark limit
$m_Q\rightarrow\infty $.
The same arguments show the suppression of the $SU(3)$
symmetry breaking
effects for vacuum condensates when the additional gluon
fields are
included.
\end{abstract}
\end{titlepage}
\vskip 0.3cm
\noindent
{\bf 1. Formulation of  the problem }
\vskip 0.3cm

It is widely believed that the QCD vacuum has a very complex
structure.
To understand this structure is a very ambitious goal which
assumes the
solution of a whole spectrum of tightly connected problems, such as
confinement,
chiral symmetry breaking phenomenon, the $ U(1)$ problem,
the $\theta$
 dependence
problem and many, many others.

A less ambitious purpose is the phenomenological study of
 the QCD
vacuum structure from the known experimental data. The most
 appropriate
analytical approach which makes contact between fundamental theory,
QCD, from the one side and observable variables from the other side,
is the QCD sum rules \cite{Shif1},\cite{Shif2}. This method has been used
extensively for about fifteen years for various purposes, mainly for
the extraction of information on hadronic properties. It is important to
stress that the enormous wealth of data referring to the low energy
hadronic physics obtained in this way,  is determined by only a few
fundamental characteristics
\be
\la \bar{q}q\ra \simeq - (0.25 GeV)^3    \nonumber
\ee
\be
\label{1}
\la \frac{\alpha_s}{\pi}G_{\mu\nu}^2\ra \simeq 1.2\cdot 10^{-2}GeV^{4}
\ee
\be
\la \bar{q}ig\sigma_{\mu\nu}G_{\mu\nu}^a\frac{\lambda^a}{2}q\ra\simeq
m_0^2\la \bar{q}q\ra ,~~~~~~ m_0^2\simeq 0.8 GeV^2 \nonumber  \nonumber
\ee
Here $q$ describes any of the light quark fields $u, d, s$ and
$G_{\mu\nu}$
is the field of colored gluons.

The value of the chiral condensate $ \la \bar{q}q\ra $ has been
 known for
a long time
\cite{Leut} and the value of the gluon condensate has been first
 extracted
from the analysis on charmonium system \cite {Shif1}. The value of
the mixed
 quark-gluon condensate has been extracted first from the analysis
of the
nucleon \cite{Ioffe}.

It is clear that the QCD vacuum is a very complicated system and a
few global
dimensional parameters (\ref{1}) give only
 some insight on the scale of vacuum
fluctuations, but do not at all exhaust all characteristics of the
vacuum
structure.

For studying of the fine aspects of the
 theory it is necessary to know the averages
of more complicated operators like $\la(\bar{q}q)^2\ra $,
 $ \la \bar{q}
G_{\mu\nu}G_{\mu\nu}q\ra $, $ \la G_{\mu\nu}^3 \ra$,
$\la G_{\mu\nu}^4\ra$, etc.
Comparing these averages with global characteristics
(\ref{1}) gives a quantitative
information of the granulated structure
of the vacuum, characterizes the correlation
between vacuum fluctuations of
large ($\sim R_{conf}$) and small ($\sim GeV^{-1}$) size,
gives some insight on the role of the
interaction of the various vacuum fluctuations, etc.
For instance, if the vacuum would be homogeneous one,
we would get more or less
an uniform distribution for gluonic $G_{\mu\nu}$
and quark $\bar{q}q$ vacuum fields.
The factorization hypothesis
 should work in this case perfectly well and predicts
that $\la G_{\mu\nu}^4\ra\simeq
{\la G_{\mu\nu}^2\ra}^2 $, $\la(\bar{q}q)^2\ra \simeq
{\la{\bar{q}q}\ra }^2$. At the same
 time, some granulated distribution of the vacuum
fluctuations, like instantons, gives a very strong
deviation from this prediction.
In particular, the instanton liquid model
\cite{Shur},\cite{Shur1} predicts the violation of
the factorization hypothesis on the level of the factor 3-10
(numerical factor depends
 on color and Lorentz structure of the operator).

Let us note that the status of this hypothesis for four-quark
condensates $ \la(\bar{q}\Gamma q)^2\ra $ (dimension six)
 has been examined in   many different ways (see \cite{Shif2}
for review).
Several channels have been studied with the special
task of detecting
deviations from the factorization for the
four-quark operators. The conclusion is as
follows: in vector and axial-vector cases
( $ \Gamma =\gamma_{\mu},~~\gamma_{\mu}\gamma_{5} $ ) deviation
from factorization formula cannot significantly
exceed $ \sim 20\% $. On the contrary, it can be shown \cite{Zhit}
 that this situation is not universal and in exotic cases with
  $ ( \Gamma =\sigma_{\mu\nu}, \gamma_{5}, 1) $ ,
a noticeable violation
of factorization does occur\footnote{Some qualitative
arguments on why it
should be so, are given in the last section.}.

The purpose of this letter is to obtain
 phenomenological, model independent
information on mixed vacuum condensates of dimension seven
$ \la \bar{q}G_{\mu\nu}G_{\mu\nu}q \ra $, which are minimal possible
operators where factorization procedure can be applied
 ($ \la \bar{q} G_{\mu\nu}G_{\mu\nu}q\ra \sim   \la \bar{q}q\ra\cdot
\la G_{\mu\nu}G_{\mu\nu}\ra $) and checked. In addition, we study
the $SU(3)$ breaking effects for these mixed condensates.
Some motivation (apart from the academic interest concerning
QCD vacuum structure) for the considering these VEVs will be
  explained in the conclusion.

Before we proceed, let us describe briefly the main idea of the
approach
which we follow in this paper. As   mentioned above, the QCD sum
rules
make contact between vacuum characteristics through their
condensate
values and  some hadronic properties. It turns out that the
heavy-light
($Q\bar{q}$) quark system is a very {\it sensitive}
 to the high dimensional
condensates like   $ \la \bar{q} G_{\mu\nu}\sigma_{\mu\nu}q\ra  $,
$ \la \bar{q} G_{\mu\nu}G_{\mu\nu}q\ra $...
Thus, some hadronic characteristics are determined by these vacuum
expectation values (VEVs).

Once this is said, the standard line of reasoning
 (very roughly) is the following. The magnitude of the
condensate  gives, on one hand,
the characteristic scale where deviation from the asymptotic
 behavior
is still under control. On the other hand the same scale
 defines some dimensional hadronic parameters which, in principle,
can be extracted by applying some fitting procedure.

We go in the {\it opposite } direction.
We are not interested in calculation
of any hadronic parameters in this paper. We are not
 calculating,
in particular, $m_{B_s}$ or $m_{B_u}$ in the limit
 $m_{Q}\rightarrow\infty$.
 We are not going to prove the power of the method by
 demonstrating that $m_{B_s} - m_{B_u} \simeq
100 MeV $ in agreement with
experimental data. {\it Instead}, we assume
that the QCD sum rules describe this system well enough.
In particular they should
explain the inequality  $ m_{B_s} > m_{B_u} $ which we
assume is still valid
in a heavy quark limit $m_{Q}\rightarrow\infty$.
Because of the {\it sensitivity} of the corresponding
sum rules (which we are going
to consider) to the higher dimensional condensates, this assumption
( $ m_{B_s} > m_{B_u} $) gives a very nontrivial restriction on the
absolute values of these VEVs. Such restrictions
are the main result of this paper.

The next,   part of the paper is devoted to the description
of the method we are going
to use. We consider the well-known $\rho - \phi$ system and find the
inequality which follows from the fact that $m_{\phi}>m_{\rho}$.
The obtained restriction is a very weak and not interesting. The only
justification for including this section to the paper is
to formulate an idea with the simplest example.
The  third   part of the paper
is its  main part where we apply the previously formulated idea to
$Q\bar{q}$ system. Some comments are given in the conclusion.

\vskip 0.3cm
\noindent
{\bf 2. The Method. }
\vskip 0.3cm

We introduce the $J^{(\phi)}_{\mu} =\bar{s}\gamma_{\mu}s$ current with the
$\phi$  meson quantum numbers and define the corresponding polarization
operator in a standard way:
\be
\label{2}
i\int e^{iqx}dx\la T\{J^{(\phi)}_{\mu}(x),~~J^{(\phi)}_{\nu}(0)\}\ra=
(q_{\mu}q_{\nu}-q^2 g_{\mu\nu}) P^{(\phi)}(Q^2),~~~~Q^2\equiv -q^2.
\ee
The calculation of the correlator for $Q^2\rightarrow\infty$ with the
power corrections taken into account is done in the standard way, and
after the Borelization procedure the sum rule takes the following form
\cite{Shif1}:
\be
\frac{1}{\pi}\int e^{-\frac{S}{M^2}}Im P^{(\phi)}(s)ds=
\frac{M^2}{4\pi^2}\{1-\frac{6m_s^2}{M^2}+
\frac{8\pi^2\la m_s\bar{s}s\ra}{M^4}
  \nonumber
\ee
\be
\label{3}
+\frac{\pi^2}{3}\frac{\la \frac{\alpha_s}
{\pi}G_{\mu\nu}^a G_{\mu\nu}^a\ra}{M^4}
-\frac{448}{81}\pi^3\alpha_s\frac{\la\bar{s}s\ra^2}{M^6}+...\}.
\ee
The matrix elements entering to eq. (\ref{3}) are discussed above
and we assumed the factorization for four-quark
operator  with ( $ \Gamma =\gamma_{\mu},~~\gamma_{\mu}\gamma_{5} $ )
\cite{Shif2}.
 In principle,
we could analyze this sum rule in a standard way
in order to reproduce  the well-known experimental data for $\phi$
meson. It turns out that it is very useful to consider a slightly
different sum rule by taking the first derivative with respect to
$M^2$. In this case, the left hand side of eq.(\ref{3}) is
proportional to $\frac{1}{\pi}\int e^{-\frac{S}{M^2}}
 Im P^{(\phi)}(s)sds$. If the $\phi$ meson would saturate
these sum rules, we would get:
\be
\label{4}
m_{\phi}^2\simeq\frac
{\frac{1}{\pi}\int e^{-\frac{S}{M^2}}  Im P^{(\phi)}(s)sds}
{\frac{1}{\pi}\int e^{-\frac{S}{M^2}}Im P^{(\phi)}(s)ds}
=\frac{\frac{M^4}{4\pi^2}\{1+ power~~ corrections\}}
{\frac{M^2}{4\pi^2}\{1+ power~~ corrections\}}.
\ee
The same procedure can be done for the $\rho$ meson with
substitution  $m_s\rightarrow m_{u,d}$
and $\la\bar{s}s\ra\rightarrow\la\bar{u}u\ra$ \footnote
{we assume an exact isotopical invariance and do not
distinguish $\la\bar{u}u\ra$
and  $\la\bar{d}d\ra$.}.
Now it is time to introduce the new parameter $R(M^2)$ defined as
\be
\label{R}
R(M^2)\equiv
\frac
{\frac{1}{\pi}\int e^{-\frac{S}{M^2}}  Im P^{(\phi)}(s)sds}
{\frac{1}{\pi}\int e^{-\frac{S}{M^2}}Im P^{(\phi)}(s)ds}
\cdot
\frac
{\frac{1}{\pi}\int e^{-\frac{S}{M^2}}  Im P^{(\rho)}(s)ds}
{\frac{1}{\pi}\int e^{-\frac{S}{M^2}}  Im P^{(\rho)}(s)sds}
\ee
If the lowest state would saturate the corresponding sum rule,
  we would get
\be
\label{5}
R(M^2)\simeq\frac{m_{\phi}^2}{m_{\rho}^2}> 1.
\ee
The  important thing for us in this paper is that
$R>1$. Intuitively it is clear and related to the fact
that  the mass of the strange meson (or meson
built from the strange quarks) is bigger than the
mass of its nonstrange partner. More important, this
inequality $R>1$ is still correct even without
the assumption on saturating   the dispersion integral
by the   lowest state.

We have to make a few very mild
assumptions
about the first $2-3$ lowest states,
 whose contribution to  the dispersion
integral is not suppressed by  the weight factor
$ e^{-\frac{S}{M^2}} $. For those states we assume that
the masses  and the corresponding coupling constants
of the strange particles bigger than their non-strange analogies
( for example, $f_{K^i}>f_{\pi^i},~~~f_{\phi^i}
>f_{\rho^i},~~~ m_{\phi^i}>m_{\rho^i},~~~i=1,2,3 ...$).
An enormous wealth of experimental data and physics
intuition tell us that it is definitely  should be true.
In this case, we are quite sure that the
inequality $ R>1 $ holds even when saturation by the
lowest resonance does not take place.

Taking into account the previous discussing and substituting
the theoretical parts of the corresponding sum rules into the
expression (\ref{R}) we get \footnote{We assume
that  the $SU(3)$ breaking effects are small enough in order to
use  an expansion with respect to this small parameter. It is clear
that in the exact $SU(3)$ limit, the ratio $R=1$.}:
\be
\label{R2}
R= 1+\Delta R_{1}+\Delta R_{2}
 \ee
\be
\label{phi}
\Delta R_{1}=
6\frac{m_s^2}{M^2}   ,~~~
\Delta R_{2}=
-16\pi^2\frac{\la m_s\bar{s}s\ra}{M^4}
-\frac{448}{27}\pi^3\alpha_s\frac{\la\bar{u}u\ra^2
-\la\bar{s}s\ra^2}{M^6}+...
\ee
We divided $\Delta R=\Delta R_{1}+\Delta R_{2}$
on purpose.
$ \Delta R_{1}$ in the above equation  denotes the perturbative
contribution
to $ \Delta R $ and $ \Delta R_{2}$ denotes the
corresponding nonperturbative part.
They have absolutely different physical meaning and have
different origin.
 It is clear that   $\Delta R_{1}$ is
related to the perturbative contribution.
The insertion (in order to calculate the polarization
operator (\ref{2}) of the nonzero quark mass ($m_s$) to the
asymptotic quark loop clearly decreases the phase volume. It
 automatically leads to the increasing of the corresponding
hadronic mass.The fact $\Delta R_1>0$ is the explicit
demonstration of it. The contributions to $\Delta R_1$
come exclusively from the small distance physics. In a sense,
$\Delta R_1$ describes some trivial kinematical factors;
there is no   deep, nonperturbative physics, involved
with $\Delta R_1$.

In the rest of this paper, we will be interested in the nonperturbative
part of the decomposition (\ref{R2}), namely $\Delta R_2$.
One may wonder whether the decomposition (\ref{R2}) is unique.
We refer to the
original paper \cite{Shif1} for a detail  discussion of
 Wilson operator expansion (OPE)
 in given context, but here we want to stress that the decomposition
(\ref{R2}) has the same status as the separation of all field fluctuations
in large and small  scales within  an (OPE) and so it
is well defined and based on   solid ground.

There is no
general theorem concerning the sign of $\Delta R_2$\footnote{ The
sign of $\Delta R_1$ is always positive by kinematics.}.
However, the physical picture of the origin of the hadron mass
assumes its nonperturbative nature. Let us note, in passing,
that in the chiral limit,   the mass of hadrons
  comes  exclusively from nonperturbative physics.

Thus, it is natural  to assume that the same, nonperturbative effects
are responsible for the mass difference of a hadron and its
partner with strange quarks. It is parametrically true
in the chiral limit $m_s\rightarrow 0$
for the $\rho - \phi$ system
 (\ref{phi}). In this case, $\Delta R_1\sim m_s^2,~~\Delta R_2\sim m_s
\gg \Delta R_1$. As we know $m_{\phi}^2-m_{\rho}^2\sim m_s$
and so, the mass difference in this case is determined
exclusively by the nonperturbative part.

In $Q\bar{q}$ system we are going to discuss in the next section,
both parts $\Delta R_1,\Delta R_2$ have the same order of magnitude
$\sim m_s$. However, we believe that the nonperturbative effects,
as well as perturbative ones, work in the {\it same} direction
in order to split the strange and nonstrange partners. All sum rules,
I am aware of, satisfy this assumption, and the formal expression
   which is given by formula
\be
\label{>}
\Delta R_2>0.
\ee
For $\phi-\rho$ system (\ref{phi}) this restriction leads
to the following inequality:
\be
\label{6}
\mid m_s\la\bar{s}s\ra\mid>
\frac{28}{27}\frac{\pi\alpha_{s}}{M_0^2}
[\la\bar{u}u\ra^2-\la\bar{s}s\ra^2],
\ee
where $M_0^2\simeq m_{\phi}^2\simeq 1GeV^2$ is the characteristic scale
when the corresponding sum rule does make sense.
Let me briefly explain this point. As usual, the analysis
of the sum rules is somewhat of an art. We could take parameter $M_0$
 to be a very large; in this case the next power corrections
could be definitely neglected, but the resulting
 inequality would be trivial and noninformative.
If we take $M_0$ to be too small, there is no    quarantee
 that a higher
power corrections will not dominate. Our choice of $M_0$, when the
main power correction is less then $20\% $ of the asymptotically
 leading term,
is the safe one in order to neglect other power corrections; usually
this parameter numerically is very close to the position
of the lowest resonance in a given channel. In particular,
for the $\rho$ meson, $M_0^2(\rho)\simeq 0.6 GeV^2$, for
$K^{\ast}$and $\phi$ it is  $M_0^2(K^{\ast})\simeq 0.8 GeV^2,~~
M_0^2(\phi)\simeq 1.0 GeV^2$ respectively. Starting  exactly
from these
points, the ``theoretical" and ``phenomenological" parts
of the sum rules fit    each other.
{}From the other side, we are  still in the  transition region,
and  very sensitive to
VEVs. To be safe enough for the
numerical estimations, we will choose the parameter $M_0$ to be
bigger than any lowest resonance involved in our analysis.
 In particular,
  for $\rho-\phi $ system we take $M_0\simeq 1 GeV^2$. For
this choice we are quite sure that the next power corrections
are small enough and  under control.

Now, a few remarks regarding ex.(\ref{6}). First of all, as it was expected,
both parts of this expression   are proportional to $m_s$.
In addition, because we know the $SU(3)$ breaking effects for
the condensate (see reviews \cite{Shur1},\cite{Rub},  \cite{Cher}
 and references
on previous papers therein),
\be
\label{7}
\Delta \equiv \frac{\la\bar{u}u\ra -\la\bar{s}s\ra }{\la\bar{u}u\ra}
\simeq 0.2,
\ee
and one can see that the  formula (\ref{6})  is trivially satisfied.
No new results can be obtained in this particular case.
However, in the next section we will derive an analogous inequality
where instead of $\la\bar{u}u\ra -\la\bar{s}s\ra $, there enters the
higher dimensional operators like
$\la\bar{u}ig\sigma_{\mu\nu}G_{\mu\nu}u\ra
-\la\bar{s}ig\sigma_{\mu\nu}G_{\mu\nu}s\ra$
or
$\la\bar{u}g^2G_{\mu\nu}G_{\mu\nu}u\ra
-\la\bar{s}g^2G_{\mu\nu}G_{\mu\nu}s\ra$
which are less familiar.
In this case we can
study the influence
of the gluonic fields on the $SU(3)$ breaking parameters
 and check the standard conjecture about weak influence of
the gluonic fields on the $SU(3)$ breaking parameters, i.e.
\be
\label{delta}
\Delta {=}?{=}\Delta_{g} {=}?{=}\Delta_{gg}
 {=}?{=}...
\ee
\be
\label{}
\Delta_{g}\equiv\frac{\la\bar{u}ig\sigma_{\mu\nu}G_{\mu\nu}u\ra
-\la\bar{s}ig\sigma_{\mu\nu}G_{\mu\nu}s\ra}
{\la\bar{u}ig\sigma_{\mu\nu}G_{\mu\nu}u\ra},~
\Delta_{gg}\equiv\frac{\la\bar{u}g^2G_{\mu\nu}G_{\mu\nu}u\ra
-\la\bar{s}g^2G_{\mu\nu}G_{\mu\nu}s\ra}
{\la\bar{u}g^2G_{\mu\nu}G_{\mu\nu}u\ra}      \nonumber
\ee
If we knew the $\Delta_{g}$ and $\Delta_{gg}$
 parameters independently, the expression
analogous to (\ref{6})
 would give us   information on absolute value of the condensate.
In the example  under consideration it leads to
\be
\label{8}
(0.25 GeV)^3\simeq\mid\la\bar{u}u\ra\mid <\frac{m_sM_0^2}{\Delta}
\frac{27}{56\pi\alpha_s} \simeq(0.55GeV)^3.
\ee
We expect an accuracy of such inequalities is about $\sim 20\% $,
as the standard accuracy of the sum rules approach. The next section
is devoted to $Q\bar{q}$ system where some more interesting
operators (instead of $\la \bar{u}u\ra$) will enter
into the game. In this case
an inequality analogous to (\ref{8}) will be less trivial and
much more interesting.

\vskip .3cm
\noindent
{ \bf 3. Heavy-light quark  system} 
\vskip 0.3cm
Introduce the $J=\bar{Q}i\gamma_{5}q$ current with the quantum numbers
of the    pseudoscalar meson and define
the corresponding polarization operator in a standard way:
\be
\label{9}
i\int e^{iqx}dx\la T\{J(x)^+,~J(0)\}\ra=P(q^2).
\ee
We use below the technique proposed in \cite{Shur2}, i.e. the
energy $e$ is used instead of $q^2 :q^2= (M_Q+e)^2,~~e\ll M_Q$.
The advantage of this technique is its simplicity; the disadvantage
is the existence of large $1/M_Q$ corrections for the real
 $D,B$-mesons.
However, as we mentioned, we are not going to extract any
information about real hadrons in this paper, we are going to
study vacuum of QCD. In this case, with assumption that
no qualitative changes will   occur in the limit $m_Q\rightarrow
\infty$, it is much easier to live in this imaginary world.
Besides that, let us note, that the first calculations of $
f_D,~f_B$ within the
sum rule approach have been done using
exactly this technique \cite{Shur2},\cite{Zhit2};
 independently the same system has been investigated
by the standard method in   ref.\cite{Aliev}.
In addition, this technique is very useful to make contact with
heavy-quark effective theory \cite{Georgi},
 see recent paper \cite{Neubert}.

A few remarks regarding new variable $e$ instead of $q^2$:
as usual, we are going to calculate the correlation function
in the nonphysical region, $e<0$ (it is analogous to $q^2<0$
in the standard method). In the limit
$E=-e\rightarrow\infty$, (but $E\ll M_Q$
 in order to use heavy quark approximation),
we can use the standard OPE and Borel transformation, defined by
operator
$\hat{B}$.
In given case $\hat{B}$ takes  the following form:
\be
\label{B}
\hat{B}\equiv \lim_{E,n\rightarrow\infty;\frac{E}{n}=M=const.}
\frac{1}{(n-1)!}E^n(-\frac{d}{dE})^n
\ee
where  the energy $E$ is positive and measured from the threshold.
The physical meaning of the Borel transformation
in this case much more clear than in the relativistic case.
Indeed, when we are passing from $E$ to $M$ we actually go
from Green function at given energy to the time Green function.
The contribution  of the
each state with energy $E_i$ to the dispersion integral  is given by
formula $e^{-\frac{E_i}{M}}$ and $\frac{1}{M}$ plays the role of
time $~~i t$. It is clear that the OPE expansion $\frac{1}{M^k}$
at $M\rightarrow\infty$ corresponds to $t\rightarrow 0$, i.e.
asymptotically small times. To study a low lying resonance we have to
calculate the correlator at small enough $M$, which corresponds to
large distance physics (large times $t$).

It is easy to check the following properties of operator $\hat{B}$:
\be
\label{B1}
\hat{B}(\frac{1}{E^k})=\frac{1}{(k-1)!}(\frac{1}{M^k}),~
\hat{B}(E^k\ln E)=(-1)^{k+1}k!M^k,~
\ee
\be
\hat{B}\frac{1}{E+E_i}=\frac{1}{M}\exp^{-\frac{E_i}{M}}\nonumber
\ee

After these few general remarks let us calculate the
nonperturbative, Borel transformed part of $ P_{np}(M)$
with strange
$ s $ quark in the current $J=\bar{Q}i\gamma_{5}s$ :
\be
\label{3.1}
P_{np}=\frac{1}{2M} \{ -\la\bar{s}s\ra+
\frac{1}{4M}\la m_s\bar{s}s\ra +\frac{1}{4M^2}\la
\bar{s}ig\sigma_{\mu\nu}G_{\mu\nu}s\ra\},~P_{p}=
\frac{1}{2M}\frac{3\cdot 2!M^3}{\pi^2}.
\ee
We removed the common factor $\frac{1}{2M}$ on purpose--
the contribution of any physical  resonance   to the dispersion
integral has the same factor $\frac{1}{2M}$  and after
Borelization takes the following form:
\be
\label{3.2}
\hat{B}\frac{1}{\pi}\int
\frac{Im P(s)ds}{s+Q^2}\rightarrow f_P^2(\frac{M_P^2}{M_Q})^2
\frac{1}{2M}\frac{1}{M_Q} e^{-\frac{E_R}{M}},
\ee
where we used the standard definition for $f_P$
\be
\la 0|\bar{Q}i\gamma_5q |P_Q\ra=f_P\frac{M_P^2}{M_Q}.
\ee
Besides that, the factor $\frac{1}{M}$  in (\ref{3.2})
is related to
Borel transformation (\ref{B}), and $\frac{1}{M_Q}$ comes
from nonrelativistic $\delta$ function :$\delta (s-M_P^2)=
\delta(2M_Q(E-E_R))=\frac{1}{2M_Q}\delta(E-E_R)$.
We included  to eq.(\ref{3.1})
the formula for perturbative, asymptotically leading part
as well. The reason to do so just to make sure that at the scale
we are going to discuss, the power corrections are much smaller
 than the leading term. But essentially, the only information
we need to know about this term is its dimension, which can be found by
pure dimensional arguments. The inequalities we are interested in and
which follow from the  our main nonperturbative assumption (\ref{>})
do not depend on the constant in front of $P_{p}$ and we will not
discuss the perturbative contribution any more.

Let us note that $P(M)$ does not depend on $M_Q$. From this
observation it is easy to reproduce the well known result
$f_{P}^2M_Q\sim 1$
of the effective heavy quark theory.

Now we are ready to repeat all steps of the previous section which
lead us to the eq.(\ref{6}) from the definition (\ref{R})and
assumption (\ref{>}).
For the given correlator (\ref{9}) the analogous formula looks
 as follows
\be
\label{3.3}
\la\bar{s}s-\bar{u}u\ra-\frac{1}{3M}\la m_s\bar{s}s\ra-
\frac{5}{12M^2}\la\bar{s}ig\sigma_{\mu\nu}G_{\mu\nu}s-
\bar{u}ig\sigma_{\mu\nu}G_{\mu\nu}u\ra >0,
\ee
which we would like to rewrite in the following form:
\be
\label{3.4}
\frac{5m_0^2}{12M}\Delta_{g}<\frac{m_s}{3}+M\Delta ,
\ee
where $m_0^2 \simeq 0.8 GeV^2$ is parameter for
mixed condensate (\ref{1});
 $m_s\simeq 0.15 GeV$ and $\Delta\simeq 0.2$ (\ref{7}).
Before we proceed with numerical estimations, let me make
a few comments on (\ref{3.3}). It is clear, that as $M\rightarrow
\infty$ this inequality is trivially satisfied because
$\la\bar{s}s-\bar{u}u\ra =\Delta|\la\bar{q}q\ra|>0$.
It becomes less trivial in the transition region.
In this case the last term, proportional to $\Delta_g$, contributes with
the opposite sign \footnote {There is no doubt in sign
$\Delta_g$\cite{Zhit3}.}. The requirement of positivity of $\Delta R_2$
impose some limitation on the absolute value of $\Delta_g$.
So, our main task is to find parameter $M_0$ where higher power
corrections can be
safely neglected, but inequality will be still interesting.
As we said above, we are not going to extend the sum rules
approach on the region
below the first resonance. In given case, from the QCD sum
rules analysis
in the heavy quark limit
(see recent papers \cite{Neubert},\cite{Shur3}, \cite{Braun})
and from experimental data (we expect that B system already
very close to heavy
quark limit), we can learn that $M_{B_u}-M_Q\simeq 0.5 GeV$ and
$M_{B_s}-M_Q\simeq 0.6 GeV$. So, our choice
\be
\label{M}
M_0\simeq 0.6 GeV
\ee
should be considered as   safe enough. Power corrections, as can
 be checked
from (\ref{3.1}) already small. With this remark in mind we will
 get the
following (rather weak, again) restriction:
\be
\label{3.5}
\Delta_g ~~ < ~~ 0.3.
\ee
 It is definitely in agreement with what we knew before \cite{Zhit3},
\cite{Dosch}.
Moreover, from the first estimate \cite{Zhit3}, we could suspect
that the $SU(3)$ breaking effects for mixed condensate are much weaker
than for the standard $\la\bar{q}q\ra$ condensate.
It would be very interesting to check
whether it is indeed true by some
independent analysis. In order to do so, we want to
consider the correlators, which include derivatives. The reason for
this choice will be clear soon.

Introduce the $J_{\mu}=\bar{Q}i\gamma_{5}\vec{i
D_{\mu}}q$ current with the quantum numbers
of the    pseudoscalar meson and define
the corresponding polarization operator in a standard way:
\be
\label{3.6}
i\int e^{iqx}dx\la T\{J_{\mu}(x)^+,~J(0)\}\ra=q_{\mu} P^{(1)}(q^2).
\ee
Here $\vec{D_{\mu}}=\vec{\partial_{\mu}}-igA^a_{\mu}\frac{\lambda^a}{2}$
 is the covariant derivative acting on the light quark.
The physical sense of the corresponding matrix element
\be
\label{3.7}
\la 0|\bar{Q}i\gamma_5\vec{iD_{\mu}}q |P_Q(q_{\mu})\ra=q_{\mu}
f_P\frac{M_P^2}{M_Q}\la x_{q}\ra.
\ee
can be understood from the
observation that    $\la x_{q}\ra$  determines the mean
  momentum fraction carried
by the light $q$ quark in the meson $P_Q$ with total momentum $q_{\mu}$.
 This information is very important for description of the
asymptotic behavior of exclusive processes; we refer to original paper
\cite{Zhit2} on this subject, but now we want to remark that
$\la x_{q}\ra \sim\frac{M}{M_Q}\ll 1$ and so, the
corresponding common factor $\frac{1}{M_Q}$ will occur
in phenomenological as well as in the theoretical parts.
It leads. of course, to the increasing   dimension of the
  $P^{(1)}(q^2)$ on one unit in comparison with $P(q^2)$ from
(\ref{9}).

As   was mentioned in \cite{Zhit2} the main power correction
to the theoretical part of (\ref{3.6}) is determined by the mixed
operator in the chiral limit. So the corresponding correlation
function is very sensitive to this operator and we have a good chance
to improve our limit (\ref{3.5}). This is the main motivation
to consider (\ref{3.6}) in this paper.

By repeating again all steps which lead us to the formula (\ref{3.1}),
we get the following  Borel transformed
expression for   $P^{(1)}$:
\be
\label{3.7}
P^{(1)}_{np}=\frac{1}{2MM_Q} \{ -\frac{4\alpha_s M}{3\pi}\la\bar{s}s\ra
(1-e^{-\frac{s_0}{M}})+
\frac{1}{4}\la m_s\bar{s}s\ra +\frac{1}{8M}\la
\bar{s}ig\sigma_{\mu\nu}G_{\mu\nu}s\ra-       \nonumber
\ee
\be
\frac{m_s}{32M^2}\la\bar{s}ig\sigma_{\mu\nu}G_{\mu\nu}s\ra\},
{}~~~~~~~P_{(p)}^1=
\frac{1}{2MM_Q}\frac{3\cdot 3!M^4}{\pi^2}.
 \ee
Few comments are in order. First of all,
as was expected, the leading
at $M\rightarrow\infty$ term proportional to
chiral condensate $\la\bar{q}q\ra$ is suppressed by the loop
correction $\sim\frac{\alpha_s}{\pi}$ and numerically small.
This fact  increases    sensitivity to the next VEV $\la
\bar{s}ig\sigma_{\mu\nu}G_{\mu\nu}s\ra$. As the second remark,
let us note the   appearance of the continuum threshold parameter
$s_0$ in the nonperturbative part $P^{(1)}_{np}$. This is the standard
parameter which accompany all sum rules, but usually it occurs
only in the
 perturbative parts. Its physical meaning can be seen
from the standard model for the spectral density as the
resonance plus continuum which
starting at $s_0$. The reason of appearing such parameter
in our case is   the high dimensions of   our currents.
The consequence   is the growth of the
spectral density  $Im P^{(1)}_{np}(s)$ with $s$. From the QCD sum rules
analysis
in the heavy quark limit
 \cite{Neubert},\cite{Shur3}, \cite{Braun},
we can learn that
\be
\label{3.8}
s_0=m_R+0.5 GeV.
\ee
It is interesting to note that this ``mnemonic rule"
works amusingly well in all known massive channels.
Anyhow, because the  condensate contribution
 comes with the small factor $\sim\frac{\alpha_s}{\pi}$,
  the influence of $s_0$ is negligible. As usual we included
to the formula (\ref{3.7})
the perturbative part as well in order to demonstrate
the smallness of the power correction at the scale $M_0\simeq 0.6 GeV$.

Let us rewrite the final inequality, analogous to (\ref{3.3})and
 which follows from   our main conjecture (\ref{>}):
\be
\label{3.9}
\frac{4M\alpha_s}{\pi}(1-e^{-\frac{s_0}{M}}-\frac{s_0}{3M}
e^{-\frac{s_0}{M}})
\la\bar{s}s-\bar{u}u\ra- \la m_s\bar{s}s\ra-
\nonumber
\ee
\be
\label{}
\frac{5}{8M}\la\bar{s}ig\sigma_{\mu\nu}G_{\mu\nu}s-
\bar{u}ig\sigma_{\mu\nu}G_{\mu\nu}u\ra
+\frac{3m_s}{16M^2}\la\bar{s}ig\sigma_{\mu\nu}G_{\mu\nu}s\ra ~~ >0.
\ee
It can be rewritten in much more clear way, using
our notations for $\Delta$ and $\Delta_g$ (\ref{delta}):
\be
\label{3.10}
\frac{5m_0^2}{8M}\Delta_{g}< m_s +\Delta
\frac{4M\alpha_s}{\pi}(1-e^{-\frac{s_0}{M}}-\frac{s_0}{3M}
e^{-\frac{s_0}{M}})
-\frac{3m_sm_0^2}{16M^2},
\ee
For numerical estimations we use the standard set of parameters
already normalized at the low normalization point:
$\mu\simeq 1GeV;~~m_s\simeq 0.15GeV;~~m_0^2\simeq 0.8 GeV^2;~~
\alpha_s(\mu)\simeq 0.34  $. Exactly
these parameters have been used in the analysis of
$\bar{q}Q$ system in the heavy quark limit \cite{Braun}
 with the special
attention on the loop corrections and we prefer
  not to change it.
With these remarks in mind we will get the following very strong
restriction on $\Delta_g$:
\be
\label{3.11}
\Delta_g~~<~~0.15.
\ee
This is in agreement with our feeling gathered from the absolutely
independent consideration \cite{Zhit3}. The relation
(\ref{3.11}) means that the gluon insertion into the chiral condensate
makes $\la\bar{s}ig\sigma_{\mu\nu}G_{\mu\nu}s\ra$ numerically
very close to $\la\bar{u}ig\sigma_{\mu\nu}G_{\mu\nu}u\ra$.

Few words on the accuracy. We expect that the next power corrections
can not exceed $10-20\%$ anyhow. Besides that, we keep only
linear $SU(3)$ breaking terms $\sim m_s$. In particular,
$\frac{\la\bar{u}u-\bar{s}s\ra}{\la\bar{s}s\ra}$ is treated
as $\Delta$ and not as $\Delta(1+\Delta)$. The term $\sim\Delta^2$
is order $m_s^2$ and should be discarded (or we must gather
all these corrections at once). We expect that these next
corrections $\sim m_s^2$ can not exceed $10-20\%$.
Finally, the reason for our belief in (\ref{3.11}) is that the
analysis \cite{Zhit1}
 of the quite different system of  Goldstone bosons,
  gives the same conclusion.

As a last remark concerning formula (\ref{3.10}), we want to stress
one more time that this is not a sum rule in the usual sense. We are
not fitting it in order to extract some information
on hadron properties. Instead, we do believe that the corresponding
sum rules reproduce mass spectrum correctly,
in particular, strange
hadron heavier than its non-strange partner. It leads
to some   relations between VEVs. We expect that the obtained
inequalities should hold in the region of variable
$M$ where sum rules are suppose to be working.

With this remark in mind and in order to get some
feeling on the high dimension condensate
$\la\bar{q}g^2G_{\mu\nu}G_{\mu\nu}q\ra $,
let us introduce the current $J_{\mu\nu}=\bar{Q}i\gamma_{5}\vec{i
D_{\mu}}\vec{iD_{\nu}}q$ and consider polarization operator
\be
\label{4.1}
i\int e^{iqx}dx\la T\{J_{\nu\mu}(x)^+,~J(0)\}\ra=q_{\mu}q_{\nu} P^{(2)}(q^2)
+g_{\mu\nu}P_{\perp}.
\ee
We will be interested in consideration of $q_{\mu}q_{\nu}$ kinematical
structure only. The corresponding matrix element
\be
\label{4.2}
\la 0|\bar{Q}i\gamma_5\vec{iD_{\mu}}\vec{iD_{\nu}}q |P_Q(q_{\mu})\ra=
q_{\nu}q_{\mu}
f_P\frac{M_P^2}{M_Q}\la x_{q}^2\ra +g_{\mu\nu}...   ~~.
\ee
determines the mean square of the longitudinal momentum
carried by light quark. It is clear that $\la x_{q}^2\ra
\sim\frac{M^2}{M_Q^2}$ and so we expect an additional suppression
$\frac{1}{M_Q}$ of the correlator in comparison with the previous case
and increasing of its dimension on one unit more. An explicit
calculation supports this expectation and the corresponding
expressions for the perturbative $P_{p}^{(2)}$ and nonperturbative
 $P_{np}^{(2)}$ parts of (\ref{4.1}) take the form
\be
\label{4.3}
P^{(2)}_{np}=\frac{1}{2MM_Q^2} \{\frac{2\alpha_s M}{3\pi}\la m_s\bar{s}s\ra
(1-e^{-\frac{s_0}{M}})+
\frac{4\pi}{81M}\la \sqrt{\alpha_s}\bar{s}s\ra^2 -\nonumber
\ee
\be
\frac{1}{24M}\la m_s
\bar{s}ig\sigma_{\mu\nu}G_{\mu\nu}s\ra
-\frac{1}{24M^2}\la O_s\ra \}, ~~~P_{p}^{(2)}=
\frac{1}{2MM_Q^2}\frac{4\cdot 4!M^5}{\pi^2}.
\ee
In this formula the operator $\la O_q\ra $ has dimension seven and defined
as follows
\be
\label{4.4}
\la O_q\ra =\la \bar{q}g^2G_{\mu\nu}G_{\mu\nu}q\ra-
\frac{1}{4}\la \bar{q}g^2\sigma_{\mu\nu}G_{\mu\nu}
\sigma_{\mu'\nu'}G_{\mu'\nu'} \ra
\ee
where as usual $G_{\mu\nu}\equiv G_{\mu\nu}^a\frac{\lambda^a}{2}$.
The factorized value of this condensate is given by
\be
\label{4.5}
\la O_q\ra_f=\la g^2 G^a_{\mu\nu}G^a_{\mu\nu}\ra\cdot\la\bar{q}q\ra
\cdot(\frac{1}{6}+\frac{1}{12})
\ee
where the coefficient $\frac{1}{6}$ comes from the factorization of
the first term (\ref{4.4}) and $\frac{1}{12}$ comes from the last
one. Let us note that the sign of $\la O \ra$ is determined uniquely
from the inequality obtained below and independently from
\cite{Zhit1}; it coincides with its factorized value (\ref{4.5}).
Thus we define
\be
\label{4.6}
\la O\ra=\la O\ra_f\cdot K ,~~~K>0
\ee
and our task now is to find some restrictions
on the coefficient of nonfactorizability $K$.
To do so we follow the standard
procedure which bring us to eqns.(\ref{3.4},\ref{3.10}):
\be
\label{4.7}
\frac{7\pi^2}{6M^2}\la\frac{\alpha_s}{\pi}G_{\mu\nu}^2\ra\cdot
K\cdot\Delta_{gg}> -m_s\cdot  \frac{32M\alpha_s}
{5\pi}(1-e^{-\frac{s_0}{M}}-\frac{s_0}{3M}
e^{-\frac{s_0}{M}})+\nonumber
\ee
\be
\frac{m_sm_0^2}{M}-
\frac{64\pi\alpha_s\Delta}{27M}|\la\bar{q}q\ra|.
\ee
With the standard input parameters it gives:
\be
\label{4.8}
K\cdot\Delta_{gg}~~>~~0.38.
\ee
As we argued above, the insertion of an additional gluon field into the
condensate $\la \bar{q}q\ra$,
 leads to the suppression of the $SU(3)$ symmetry
breaking effects in the obtained mixed condensate $\la \bar{q}Gq\ra$
of approximately one quarter: $ {\Delta_g}/{\Delta}\simeq 0.75$.
We expect that the same suppression holds for   each of the next
gluon insertions. Thus, for $0.75 \Delta_g<\Delta_{gg}<\Delta_g$
we obtain:
\be
\label{4.9}
K~~>~~2.5 (for \Delta_{gg}=\Delta_g);~~K~~>~~3.4 (for \Delta_{gg}=
0.75\Delta_g),
\ee
which is our main result. The independent analysis of the different
system (Goldstone bosons, not $\bar{Q}q$ mesons) supports this
result and demonstrates the violation of factorization by a factor 3.
\vskip .3cm
\noindent
{\bf 4.Conclusion}
\vskip 0.3cm
Thus we have found two qualitative phenomena from the analysis of
the $\bar{Q}q$
system (very sensitive to vacuum structure)
in the heavy quark limit $M_Q\rightarrow\infty$:

$\bullet$
$SU(3)$ breaking effects are being suppressed with the insertion
of gluon fields into condensate.
In particular, we expect that
the following link of inequalities is correct
$\Delta>\Delta_g>\Delta_{gg}>...$
I do not know whether any
QCD vacuum models can  (at least) qualitatively describe
this behavior.

$\bullet$
The factorization hypothesis does not work for mixed, dimension seven
operator $\la \bar{q}GGq\ra$. It is not a big surprise, because
we faced the analogous phenomenon early \cite{Zhit} for the
four-fermion condensates with an "exotic"\footnote{
This term has been borrowed from  the Shifman Introduction
to the book \cite{Shif2}}   Lorentz
structure.

In conclusion, I want to give some qualitative remark which
elucidates the reason why some VEVs must violate factorization.
First of all, let me remind you that in some special cases
the QCD sum rules do not work in principle. In particular,
 in the pseudoscalar  quark channel
\be
\label{a}
i\int e^{iqx}dx\la T\{J^{(\pi)} (x),~J^{(\pi)} (0)\}\ra=
  P^{(\pi)}(Q^2),~Q^2\equiv -q^2,~J=\bar{u}i\gamma_5d
\ee
the standard sum rule
can not reproduce the residue $\la 0| J|\pi\ra=f_{\pi}
\cdot 2GeV$ which is known
exactly from PCAC, and which has much
bigger  scale than QCD sum rule can provide.
 The reason for that became clear
very soon after QCD sum rules were invented and related to the fact
that the limited dynamical information encoded by means
of the different VEVs is not sufficient for describing
the vacuum $0^{\pm}$ channels. So called direct instanton
contributions \cite{Novik} play a decisive role
in these channels. Let us call these currents  as ``exotic" ones,
to emphasis their difference in comparison with the ``normal" currents.
The scale provided by QCD sum rules analysis in the ``normal" cases
does describe all residues correctly.

Now let us consider any nondiagonal correlator, constructed
from the ``normal" and ``exotic" currents. On the one hand,
  there are no direct instanton contributions; on the other
hand, the ``exotic" matrix element $\la 0| J|\pi\ra =
f_{\pi}2GeV $ is
several times larger than the normal scale $f_{\pi}\cdot m_{\rho}$,
which the sum rule can provide. Thus, the information about
big scale connected with contributions of direct instantons,
clearly filters through into $Im P(s)$ via the residue
$r_{\pi}\sim \la| J^{exotic}|\pi\ra\la\pi| J^{normal}| \ra$.
The question arises, in which way can one guarantee a large scale
of the quantity $r_{\pi}$, if the corresponding sum rules
do not allow direct instanton contribution? From our point of
view the answer is that some VEVs which enter into the corresponding
correlator become  {\it numerically} large. Precisely in such a way
{\it(through enhancement of the vacuum condensates)} the information
of large value of the scale {\it filters} through into correlators,
which do not allow direct instanton contributions.

$\bullet$
Few words on applications obtained results. First of all, as was explained
in the text, these mixed VEVs determine the momentum distribution between
$q$ and $Q$ quarks in the heavy-light system. In any calculation
with $\bar{q}Q$ system involved, this information is very useful, see e.g.
\cite{Griffin}.

As a second remark, I would like to note that the transverse quark distribution
 in  the light
mesons is determined mainly by high-dimensional condensates studied
above, see \cite{Zhit1} for a more details. As is known, the transverse
size dependence plays a key role in color transperency physics \cite{Brod}.
 \vskip .3cm
\noindent
{\bf 5.Acknowledgements}
\vskip 0.3cm
 
The present analysis
and the introducing of the
parameter $R$ (\ref{R}) was motivated by paper Ben Grinstein \cite{Grinst}
where he considered the double ratio
$R_1=\frac{f_{B_s}/f_{B}}{f_{D_s}/f_{D}}$. I want to thank   him for
 conversations. This work is supported by the Texas National Research
Laboratory Commission under  grant \# 528428.


\begin{thebibliography}{xx}

\bibitem{Shif1}
M.A.Shifman, A.I.Vainshtein and V.I.Zakharov \\
   Nucl.Phys.  {\bf B147}, (1979)385,448,519.

\bibitem{Shif2}
M.A.Shifman, Vacuum Structure and QCD Sum Rules, North-Holland,1992.
\bibitem{Leut}
H.Leutwyler, Phys.Lett.{\bf B48},(1974),45.
\bibitem{Ioffe} V.M.Belyaev and B.L.Ioffe
Sov.Phys.JETP{\bf 56},(1982),493.

\bibitem{Shur} E.Shuryak
Nucl. Phys.{\bf B203},(1982),93,117.\\
D.Diakonov and V.Petrov, Nucl.Phys. {\bf B245},(1984),259.
\bibitem{Shur1} E.Shuryak,
The QCD Vacuum,Hadrons and The Superdense Matter,
World Scientific Publishing Co.,1988.
\bibitem{Zhit} A.Zhitnitsky,Sov.J.
Nucl. Phys. {\bf 41},(1985),513.

\bibitem{Rub} L.J.Reinders, H.Rubinstein and S.Yasaki,
Phys.Rep. {\bf 115},(1984),151.
\bibitem{Cher} V.L.Chernyak and A.R.Zhitnitsky,
Phys.Rep. {\bf 112},(1984),173-318.
\bibitem{Shur2} E.~Shuryak, Nucl.Phys. {\bf B198},(1982),83.

\bibitem{Zhit2}
V.L.Chernyak, A.Zhitnitsky, I.Zhitnitsky
Sov.J.Nucl.Phys.{\bf 38}, (1983),775; see also review article\cite{Cher}.

\bibitem{Aliev}T.Aliev and V.Eletsky,
Sov.J.Nucl.Phys. {\bf 38},(1983),936.
\bibitem{Georgi} H.Georgi, Phys. Lett. {\bf 240},(1990),447.
\bibitem{Neubert} M.Neubert,
Phys.Rev.. {\bf D 45},(1992)2451;
\bibitem{Zhit3} V.Khatsimovsky,I.B.Khriplovich and A.Zhitnitsky,
 Z.Phys. {\bf C36},  (1987), 455.
The main task of this paper was the calculation
of the matrix element $\la N|\bar{s}s|N\ra$
and the neutron dipole moment in the
Weinberg model of CP violation. This problem is reduced to
 the evaluation of the  vacuum characteristic $\Delta_g$.
In the appendix of the paper, we made a  rough estimation of this VEV:
 $\Delta_g\simeq
0.15$. A more careful
analysis \cite{Dosch},
exploiting  the same idea, confirmed this estimate.
\bibitem{Braun} E.Bagan et al,Phys. Lett. {\bf B278},(1992),457.
\bibitem{Shur3} V.Eletskii, E.Shuryak, Phys.Lett {\bf B276},(1992),191.

\bibitem{Dosch}M.Beneke and H.G.Dosch,
Phys. Lett. {\bf B 284}, (1992),116.  
\bibitem{Zhit1} A.Zhitnitsky,
Goldstone Bosons, QCD Vacuum
Structure and Color Transperency Physics, Preprint SMU-HEP-93-25,
 Dallas, November,1993.
\bibitem{Novik}
V. Novikov et al.,
Nucl. Phys. {\bf B191},(1981),301.
\bibitem{Griffin} P. Griffin, M.McGutgan, M.Massip, Florida Preprint
UFIFT-HEP-93-25, November, 1993.
\bibitem{Brod} S.Brodsky and A. Mueller, Phys. Lett.{\bf B 206}, (1988),685.
\bibitem{Grinst} B.Grinstein, SSCL-Preprint-492,hep-ph/9308226

\end{thebibliography}
\end{document}